\documentclass[unsortedaddress,superscriptaddress,prb,twocolumn]{revtex4-2}
\usepackage{hyperref}
\usepackage{epsfig}
\usepackage{graphicx}
\usepackage{subfigure}
\usepackage{latexsym}
\usepackage{color}
\usepackage{fullpage}
\usepackage{dcolumn}
\usepackage{bm}
\usepackage[normalem]{ulem}
\usepackage{units}
\usepackage{amsmath}
\usepackage[paperwidth=210mm,paperheight=297mm,centering,hmargin=2cm,vmargin=2.3cm]{geometry}

\begin{document}


\title{Epitaxial two-dimensional membranes under intrinsic and extrinsic strains}

\author{Nicolas Rougemaille}
\affiliation{Universit\'{e} Grenoble Alpes, CNRS, Institut NEEL, Grenoble INP, 38000 Grenoble, France}
\author{Johann Coraux}
\email{johann.coraux@neel.cnrs.fr}
\affiliation{Universit\'{e} Grenoble Alpes, CNRS, Institut NEEL, Grenoble INP, 38000 Grenoble, France}

\begin{abstract}
Two-dimensional (2D) materials naturally form moir\'{e} patterns with other crystalline layers, such as other 2D material or the surface of a substrate. These patterns add a nanoscale characteristic length in the form of a superlattice: the moir\'{e} wavelength. Understanding the origin and characteristics of these patterns is crucial to design/interpret moir\'{e}-induced physical properties. Here, we use a mixed continuum mechanics + atomistic modeling to study two experimentally relevant epitaxial 2D materials ---graphene on Ir(111) and MoS$_2$ on Au(111)--- under extrinsic and intrinsic strain. We consider three different scenarios affecting substantially the lattice constant of the 2D materials, the wavelength and corrugation of the moir\'{e} pattern. (i) Under the influence of the interaction with the substrate, bending energy produces non trivial variations of the moir\'{e} properties, even when the strain is small; (ii) When locked on a progressively strained substrate via the valleys of the moir\'{e}, the membranes' nanorippling amplitude goes through several jumps related to relatively smaller jumps in the interatomic distance of the 2D materials; (iii) Finally, increasing the zero-deformation value of this interatomic distance (possibly controlable with temperature or illumination in experiments) the moir\'{e} wavelength can either increase or decrease.
\end{abstract}

\maketitle

\section{Introduction}

Moir\'{e} patterns in two-dimensional (2D) materials like graphene, boron nitride and transition metal chalcogenides are a key ingredient allowing to enrich the properties of these flexible, atomically thin membranes \cite{Andrei2020,Shabani2021,Yasuda2021,Xu2022}. Deformations, both at the atomic scale and at the larger scale of moir\'{e} nanorippled patterns, are relevant here \cite{Deng2016,Dai2019}. This is actively studied in twisted bilayers of 2D materials with diverse degrees of control on the geometry of the moir\'{e} patterns \cite{Qiao2018,Pena2023,Kapfer2023}. When the layers have different lattice parameters, which can be adjusted by strain imposed to one of the layers, considerable modifications of the electronic properties are expected \cite{Khatibi2019}, giving rise to the concept of moir\'{e} gravity within a cosmology viewpoint, and even to gauge fields experienced by the electronic states in presence of out-of-plane deformations \cite{Parhizkar2022}. The role of the moir\'{e} patterns in bilayer systems is also possibly prominent concerning the relative interfacial sliding between the two layers, bringing peculiar frictional behaviour \cite{Wang2019,Wang2019b} and possibly superlubricity \cite{Yang2021}.

\begin{figure}[!hbt]
\begin{center}
\includegraphics[width=8cm]{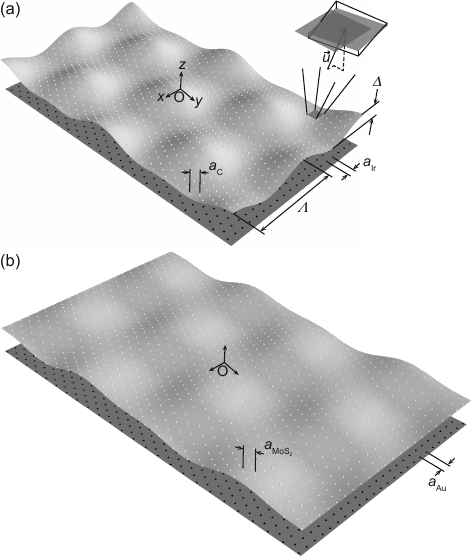}
\caption{\label{fig1} Graphene (a) and MoS$_2$ (b) single layers on their crystalline substrates, each with a distinct nano-undulation pattern (cosine and sine respectively). The strength and spatial variations of the interaction potential between the substrate (lattice constant, $a_\mathrm{s}$) and the 2D material, which depend on the relative atomic coincidences, together with the elastic energy stored in the 2D membrane, set the moir\'{e} lattice's structural parameters (wavelength $\Lambda$, rippling amplitude $\Delta$, membrane lattice constant $a_\mathrm{C,MoS_2}$). The variations of these three parameters with $a_\mathrm{s}$ and the zero-deformation values of $a_\mathrm{C,MoS_2}$, $a^0_\mathrm{C,MoS_2}$, are studied here. The top cartoon illustrates a possible local deformation of the membrane's surface (first in-plane, between the two gray-shaded areas, then out-of-plane, from the gray-shaded area to the black frame).}
\end{center}
\end{figure}

Epitaxial 2D materials, i.e. 2D materials that have been synthesized onto crystalline substrates, are also characterized by moir\'{e} nanorippled patterns \cite{Grant1970,Land1992,Forbeaux1998,Corso2004,Sorensen2014}. These patterns introduce a spatial modulation of the 2D material's electronic properties \cite{VazquezdeParga2008,Schulz2014,Krane2018} and superlattice effects, including replicas of electronic bands \cite{Bostwick2007,Pletikosic2009} and mini-bandgaps \cite{Pletikosic2009}, whose positions depend on the moir\'{e} structure, and its wavelength $\Lambda$ in particular. Understanding the influence of metal / 2D material moir\'{e}s is a lively field of research, which has for instance revealed graphene phonon enhancement via top-deposited molecules \cite{Wu2023}, rotational ordering of incommensurate confined monolayers in presence of competing lengthscales \cite{Lisi2022}, a modified interaction between Kondo impurities \cite{Trishin2023} or nanoscale charge modulations \cite{Reidy2021}. Knowledge on $\Lambda$, and of the nanorippling height modulation $\Delta$, is hence crucial to understand/control moir\'{e}-related properties. This motivated us to understand how $\Lambda,\Delta$ are selected by the competition between the elastic, membrane-like properties of epitaxial 2D materials on one hand, and the interaction with their substrate on the other hand \cite{Artaud2022}. It was accordingly possible to explain a mechanical instability in epitaxial graphene on Ir(111), wrinkling, occurring as the strength of the graphene-substrate interaction changes \cite{Vlaic2018} and not only due to a mismatch of thermal expansion coefficients in the substrate and the 2D material \cite{Chae2009,NDiaye2009}.

A more obvious way to modify the moir\'{e} topography is to induce mechanical strain within the 2D material, i.e. a compression or extension of interatomic bonds. This is accompanied with an elastic energy penalty, which can be relieved by either further rippling or unrippling. In principle this can be achieved in several practical situations: applying mechanical stress to the substrate, varying the temperature to take benefit of the different lattice compression/expansion in the substrate and 2D material \cite{Chae2009,NDiaye2009}, or illuminating the 2D material \cite{Haleoot2017,Liu2022}.

Here, we explore strain-induced changes of the moir\'{e} topography using a mixed continuum mechanics + atomistic model \cite{Artaud2022}. With this method it is possible to address systems comprising several 100 atoms, with no need to impose commensurability between the lattices of the substrate and of the 2D material as in most density functional theory (DFT) approaches treating the substrate in a realistic way. It is even possible, to the expense of relatively large computational times though, to tune different control parameters for the structure, and strain in particular, which cannot be tuned freely with numerical frameworks requiring commensurability. More specifically, we consider two prototypical systems, epitaxial graphene on Ir(111) and MoS$_2$ on Au(111) that are widely studied by experimentalists (see e.g. Refs.~\citenum{Reidy2021,Lisi2022,Wu2023,Trishin2023}). We first determine the equilibrium configurations of the moir\'{e} patterns, in terms of $\Lambda$, $\Delta$ and lattice parameter ($a$, within the 2D membranes), as the lattice constant of the substrate is varied -- an experimentally relevant situation of heteroepitaxial stress. We find that bending effects play an important role in the precise value of $\Lambda$. Next we revisit this question by imposing a constraint to the membrane, namely by locking it on its substrate at the valleys of the moir\'{e}, where the atomic lattices of the substrate and of the 2D materials are in registry. This is, once more, an experimentally relevant situation, where the 2D membrane is prevented to slide on its substrate. We find that the 2D materials experiences a series of mechanical transformations, whereby $\Delta$ abruptly varies for certain substrate strain values. Finally, we study the influence of the zero-deformation value of the lattice constant within the 2D material. This quantity, which can vary with temperature or illumination in certain 2D materials, is found to influence the moir\'{e} topography in a different way substrate strain does.

\section{Methods}

\begin{figure}
\begin{center}
\includegraphics[width=7.063cm]{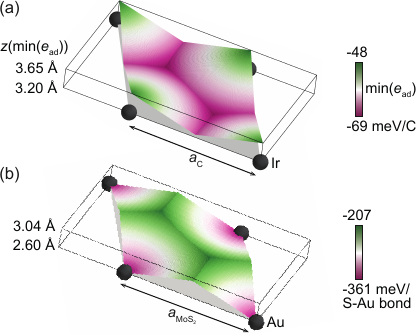}
\caption{\label{fig1b} Three-dimensional coloured surface representing the minimum of the graphene/Ir(111) (a) and MoS$_2$/Au(111) (b) adhesion energies as function of the $x,y$ position of a membrane atom with respect to the substrate atoms' in-plane positions.}
\end{center}
\end{figure}

Our model considers the 2D materials as flexible and deformable membranes on one hand, within a continuum mechanics viewpoint, and on the other hand as atomic lattices that interact with their crystalline substrates according to the relative positions of the atoms (Fig.~\ref{fig1}). The membranes' mechanical properties are treated within the thin plate theory, which assumes phenomenological elastic constants. The adhesion with the substrate is modeled starting from existing density functional theory (DFT) calculations of the interaction potential \textit{versus} the distance between the substrate and 2D material. The details of this potential are varying with the relative in-plane positions of the atoms in the substrate and the 2D material (see Fig.~\ref{fig1b}). A parametric analytic form of the potential is then sought for, with parameters varying with the relative atomic positions. Here, any interfacial lattice mismatch produces a periodic spatial variation of these relative positions, leading to a periodic lattice (a superlattice) of (quasi)coincidences.

This superlattice is the well-known moir\'{e} pattern, with wavelength $\Lambda$ (its period), and the effect of the adhesion potential is to create a nanoripple pattern with the same wavelength, and amplitude $\Delta$. Of course, $\Lambda$ and $\Delta$ depend on the relative values of the lattice constants of the substrate, $a_\mathrm{s}$, and of the 2D material, $a$, and on the twist angle (which we disregard here, but was considered in our earlier works \cite{Artaud2022}). Note that $a$ is really the lattice constant within the membrane's local surface, and not its projection along the substrate's surface (which we assume as flat, i.e. unperturbed by the presence of the 2D material). Choosing a simple, yet realistic 2D shape for the membranes based on sine or cosine functions, it is then possible to numerically compute the total energy of the system, including elastic and adhesion contributions, which may or may not compete. This is precisely what we have done in our recent work \cite{Artaud2022}, in the spirit of previous works on bilayers of 2D materials \cite{Nam2017,Enaldiev2020}, but taking into account out-of-plane deformations which can play important roles in the case of epitaxial 2D materials. Although our model can treat strain inhomogeneities, for simplicity we assume a uniform $a$ in the following. This assumption is reasonable in the present case of small-$\Lambda$ ($<$ 10~nm) moir\'{e}s, for which the variation of interatomic distances is expectedly very small across a moir\'{e} unit cell \cite{Gao2017}.

The system's elastic energy within a moir\'{e} unit cell writes as a continuous sum,

\begin{equation}
\begin{split}
E_\mathrm{el} =& \frac{1}{2} \iint \big(\lambda (\varepsilon_{xx}+\varepsilon_{yy})^2 \\
&+ 2 \mu (\varepsilon_{xx}^2+\varepsilon_{yy}^2+2\varepsilon_{xy}^2)\big) dxdy \\
&+ \frac{\kappa}{2} \iint \big( (\partial_{x}\partial_{x} u_z
+ \partial_y\partial_y u_z)^2 + 2(1 - \nu)\\ 
&\times((\partial_x\partial_y u_z)^2 - \partial_x\partial_x u_z \partial_y\partial_y u_z  ) \big) dxdy,
\end{split}
\label{eq1}
\end{equation}

\noindent with $\lambda,\mu$ the Lam\'{e} coefficients, $\kappa$ the bending constant and $\nu$ the Poisson ratio. Different values for these constants are summarized in Table~\ref{tab1} for graphene and MoS$_2$ (we disregard their possible variations with the crystallographic direction \cite{Gao2009,Georgantzinos2010,Jiang2015}). We do not consider the effect of temperature, which may change the values of the mechanical constants, in a so-far unknown way in the case of on-substrate 2D materials. Whenever possible in our calculations we use experimental values for the epitaxial systems. When they do not exist, we use experimental values for isolated single layers, or if no other value is available, those for the corresponding three-dimensional material. Equation~\ref{eq1} combines the components of the strain tensor, $\varepsilon_{\alpha\beta}$ with $\alpha,\beta=x,y$, and the spatial derivatives of the $z$ component of the displacement vector, $\vec{u}$ (see Fig.~\ref{fig1}). The former are defined as $\varepsilon_{\alpha\beta}=1/2(\partial_\beta u_\alpha + \partial_\alpha u_\beta + \sum_{\tau=x,y,z} \partial_\alpha u_\tau \partial_\beta u_\tau )$, with $\partial_{\alpha,\beta}=\partial/\partial_{\alpha,\beta}$. Importantly, all spatial derivatives of $\vec{u}$ implicitly involve the 2D material's lattice constant and its deviation from its value in the absence of deformation of the atomic bonds, the zero-deformation value $a^0$ whose influence will be studied in Section~\ref{sec:internalstrain}.

\begin{table*}
\caption{\label{tab1} Elastic constants of free-standing and on-substrate graphene and MoS$_2$ (Lam\'{e} parameters $\lambda$ and $\mu$, Poisson's ratio $\nu$, bending modulus $\kappa$), and structural parameters of the 2D materials and of the substrates [lattice constants (l.c.) and thermal expansion coefficients ($\alpha$)], from experiments and simulations (*) performed within different theoretical frameworks.}
\begin{ruledtabular}
\begin{tabular}{lllllll} 

 & Graphene & Graphene/Ir & MoS$_2$ & MoS$_2$/Au(111) & Ir(111) & Au(111)  \\

\hline

$\lambda$ (eV/\AA$^2$) & 3.60 \cite{Lee2008} & 4.21 \cite{Politano2015} & - & - &   &   \\ 
                       & 2.41*,3.29*\cite{Zakharchenko2009,Kudin2001} & - & 3.29* \cite{Enaldiev2020} & - &   &   \\ 

$\mu$ (eV/\AA$^2$) & 9.12 \cite{Lee2008}  & 8.98 \cite{Politano2015} & - & - &   &   \\ 
                   & 9.95*,9.38* \cite{Zakharchenko2009,Kudin2001} & - & 3.6* \cite{Enaldiev2020} & - &   &   \\ 

$\nu$ & - & 0.19 \cite{Politano2015} & - & - &  & \\ 
      & 0.12*,0.149* \cite{Zakharchenko2009,Kudin2001} & - & 0.29* \cite{Cooper2013} &  - &   &   \\ 

$\kappa$ (eV) & 1.17\footnotemark[1] \cite{Nicklow1972} & - & 6.0 \cite{Jiang2023} & - &   &   \\ 
              & 1.1*,1.41* \cite{Fasolino2007,Lu2009}  & - & 9.61* \cite{Jiang2013,Xiong2016} &  - &   &   \\ 

l.c. (\AA) & 2.462\footnotemark[1] \cite{Dwight1972} & 2.455,2.462 \cite{Blanc2012} & 3.167\footnotemark[1] \cite{ElMahalawy1976} & 3.163 \cite{Sant2020} & 2.715 & 2.883 \\ 
          & 2.457* \cite{Zakharchenko2009} & - & - & - & - & - \\ 

$\alpha$ (10$^{-6}$~K$^{-1}$) & [-8,-5.5] \cite{Bao2009,Singh2010,Yoon2011,Pan2012} &   & [6.7,7.3] \cite{Huang2014,Gan2016,Wang2015}  &   & 6.47 \cite{Arblaster2010} & 14.00 \cite{Dutta1963} \\
                              & -3.6* \cite{Mounet2005}  &   &  7.2* \cite{Sevik2014} &   &   &  \\

\end{tabular}
\end{ruledtabular}
\footnotetext[1]{Values for the bulk material.}
\end{table*}

We write the adhesion energy as a sum over all $N$ atoms of the membrane comprised within the moir\'{e} unit cell,

\begin{equation}
E_\mathrm{ad} = \sum_{i\in[1,N]} e_\mathrm{ad} \left(x_i,y_i,u_{z,i}\right),
\label{eq2}
\end{equation}

\noindent with $x_i,y_i$ the in-plane position of the 2D material's atoms, and $e_\mathrm{ad}$ in the form of a Morse potential, $p\times(1-e^{-q\times(u_{z,i}-r)})^2+s$. The parameters $p,q,r,s,$ depend on $x_i,y_i$ and are determined by seeking best fit of the Morse potential with the DFT data. The minimum of $e_\mathrm{ad}$ and the corresponding distance between the 2D material and its substrate are represented for epitaxial graphene on Ir(111) and MoS$_2$ on Au(111) in three-dimensional surface plots in Fig.~\ref{fig1b}. For both systems the parameters have been determined previously (see Ref~\citenum{Artaud2022} for details about the DFT data and Morse parametrization).

In the present work, using a home-made multi-processor python code running on a  10-core Intel Xeon W-2150B 3.00 GHz CPU, we have sought for low-total-energy (i.e. $E_\mathrm{el}+E_\mathrm{ad}$) configurations among extended sets of configurations, where $\Lambda$, $\Delta$, $a$, $a^0$ and $a_\mathrm{s}$ were varied, in some cases imposing specific constraints to the system, as will be the case in Section~\ref{sec:noslide}. We considered height variations with the position $\vec{r}$ in the form $\Delta/3+\Delta/9(\cos\vec{k}_1\cdot\vec{r}+\cos\vec{k}_2\cdot\vec{r}+\cos\vec{k}_3\cdot\vec{r})$ for graphene and $\Delta/2+\Delta\sqrt{3}/3(\sin\vec{k}_1\cdot\vec{r}+\sin\vec{k}_2\cdot\vec{r}+\sin\vec{k}_3\cdot\vec{r})$ for MoS$_2$, with $\vec{k}_1,\vec{k}_2,\vec{k}_3$ three vectors forming 120$^\circ$ angles and having a norm of $4\pi/(\Lambda\sqrt{3})$.

\section{Membrane deformations versus substrate strain}

In-plane compressive or tensile strain in the substrate, i.e. a variation of $a_\mathrm{s}$ with respect to a reference value (usually, the bulk, equilibrium value at a given temperature), is one of the structural parameters that can be controlled experimentally. This is the case for example when graphene, grown on a substrate at elevated temperature, wrinkles upon cooling down due to the distinct thermal expansion (Table~\ref{tab1}) in the two materials \cite{Chae2009,NDiaye2009}. Another possible implementation of substrate strain could be a mechanical stress applied macroscopically to the substrate. In all these situations, substrate strain can reach typically 1\%, rarely more.

An interesting question concerns the equivalence of a positive (negative) substrate strain with a negative (positive) strain within the 2D material. In our framework, where the lattice constant within the 2D material is assumed spatially uniform, the two situations are equivalent. A follow-up question is whether experimentally, it is possible to directly control the strain of an epitaxial 2D material, i.e. to stress it independently from the substrate, although this appears challenging. Another way to generate strain in the 2D material, probably more relevant experimentally, will be discussed in Section~\ref{sec:internalstrain}.

\begin{figure}[!b]
\begin{center}
\includegraphics[width=8cm]{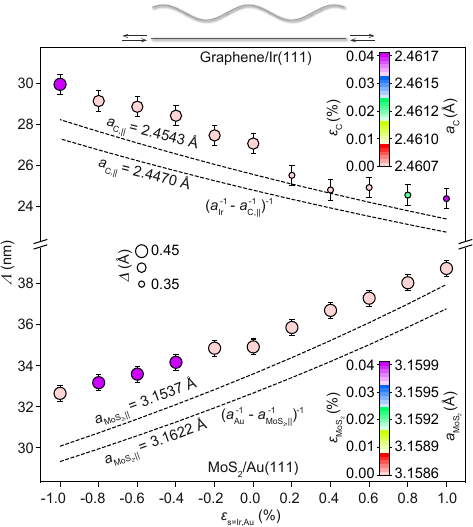}
\caption{\label{fig2} Influence of in-plane biaxial deformation of the substrate lattice. The variations of $\Lambda$, $\Delta$ and $a_\mathrm{C,MoS_2}$ in the graphene and MoS$_2$ moir\'{e} lattices are represented as function of the substrate's in-plane deformation $\varepsilon_\mathrm{s}$, with respect to the room-temperature equilibrium lattice constant (Ir(111): 2.715~\AA; Au(111): 2.883~\AA). The $a_\mathrm{C,MoS_2}$ values are coded with the color of the symbols; the $\Delta$ values are represented with the size of the disk-symbols. Dotted lines show values calculated disregarding the influence of nanorippling, for two sets of room-temperature experimental values of the in-plane projection of the 2D material's lattice constant [values from Refs.~\citenum{Blanc2012} and \citenum{Sant2020} for graphene/Ir(111) and MoS$_2$/Au(111) respectively].}
\end{center}
\end{figure}

For both graphene/Ir(111) and MoS$_2$/Au(111), we explored 7,000 configurations with varying values of $\lbrace\Lambda,\Delta,a\rbrace$. For each $a$ value, we determined the minimum of the total-energy 2D surface \textit{versus} $\Lambda$ and $\Delta$, and finally identified the $a$ value yielding global energy minimum (and the corresponding $\Lambda,\Delta$ values) \cite{Artaud2022}. The procedure was repeated for 11 substrate strain ($\varepsilon_\mathrm{s}$) values.

The result is represented in Fig.~\ref{fig2} for the two systems. As expected, the moir\'{e} wavelength $\Lambda$ increases as the lattice mismatch between the 2D material and its substrate decreases, i.e. towards increasing negative $\varepsilon_\mathrm{s}$ for graphene/Ir(111) ($a_\mathrm{C}$ remains smaller than $a_\mathrm{Ir}$) and positive $\varepsilon_\mathrm{s}$ for MoS$_2$/Au(111) ($a_\mathrm{MoS_2}$ remains larger than $a_\mathrm{Au}$).

Qualitatively, the general trend is similar to what a basic calculation of $\Lambda=1/|1/a_\parallel-1/a_\mathrm{s}|$ \cite{Artaud2016}, neglecting any nanorippling, predicts (dotted curves in Fig.~\ref{fig2}, calculated with $a_\parallel$ values of the in-plane projections of $a$ measured in high-resolution diffraction experiments \cite{Blanc2012,Sant2020}). There are, however, significant deviations to this purely 2D viewpoint, as our model systematically predicts larger $\Lambda$ values. This confirms one of the key conclusions of our previous work \cite{Artaud2022}, namely that the accumulation of bending energy penalty in the membrane is mitigated by increasing $\Lambda$. This is here put in evidence studying the effet of substrate strain, while only the influence of $e_\mathrm{ad}$ was addressed previously.

Beyond the sign of the $\Lambda$ \textit{versus} $\varepsilon_\mathrm{s}$ slope, whose origin is rather trivial, there are noticeable differences between graphene/Ir(111) and MoS$_2$/Au(111). The former not only adjusts its $\Lambda$ value, but also its nanorippling amplitude $\Delta$ (0.35-0.45~\AA) and its lattice constant (0-0.04\%). The variations of $\Delta$ are inexistant for MoS$_2$/Au(111).

Interpreting these different behaviours is not straightforward. Obviously, pure 2D strains are more costly for graphene than for MoS$_2$ (stronger interatomic bonds, larger Lam\'{e} parameters, cf. Table~\ref{tab1}), while bending graphene is less costly (graphene is thinner, and while Poisson's ratio is smaller, $\kappa$ is much smaller, see Eq.~\ref{eq1} and Table~\ref{tab1}). This does not, however, explain the outcome of our calculations. More relevant here are the differences in the values of the minimum of $e_\mathrm{ad}$, at different sites of the moir\'{e}. They are significantly larger in the case of MoS$_2$ \cite{Artaud2022} (Fig.~\ref{fig1b}). Hence, for MoS$_2$ there is stronger tendency to adopt a well-defined value of $\Delta$ (0.45~\AA, no variations in the $\varepsilon_\mathrm{s}$ range explored): there is a strong, substrate-induced selection of the nanorippling. In contrast, in the case of graphene $\Delta$ can vary.

For positive $\varepsilon_\mathrm{s}$ values, where the graphene/Ir lattice mismatch increases, $\Lambda$ decreases, tending to increase the bending energy penalty. One way for the membrane to circumvent this cost is to decrease $\Delta$. Beyond a certain $\varepsilon_\mathrm{s}$ value, further decreasing $\Delta$ becomes too costly ($e_\mathrm{ad}$ penalty) however, so $a$ increases to limit the lattice mismatch, hence an excessive decrease of $\Lambda$ (and bending energy penalty). The same line of thoughts explains the increase of $a$ for MoS$_2$ at increasing negative $\varepsilon_\mathrm{s}$ values.

\section{Low-energy configurations with substrate strain and without interfacial sliding}
\label{sec:noslide}

Until now, we have sought for the equilibrium energy configuration of the system, imposing no constraint on the way the 2D membrane can change its topography and the compression/extension of its interatomic bonds. In other words, we have assumed a total absence of energy barriers between different membrane configurations on their substrate. In an experiment, where a control parameter varies (more-or-less) continuously, this assumption might be too strong. One can for instance think of the Peierls-Nabarro barrier \cite{Peierls1940,Nabarro1947} related to the lattices' discreteness, which opposes to the motion of interfacial dislocations -- actually, a moir\'{e} lattice is nothing else than a 2D array of dislocations \cite{Pochet2017}. Experimental evidence exists that $\Lambda$ hardly varies over several 100~K below graphene's growth temperature on Ir(111) \cite{Hattab2012,Jean2013}, suggesting that the graphene lattice cannot glide over the substrate lattice while the latter compresses, despite a mismatch of thermal expansion coefficients. Instead, the two lattices are `locked-in', due to the weak but non-zero covalent-like contribution to the adhesion energy in graphene/Ir \cite{Busse2011} (also present in MoS$_2$/Au \cite{Silva2022}).

The `lock-in' scenario should apply to various 2D material / substrate systems, as long as the covalent-like contribution to the adhesion energy is non-negligible. This contribution translates into variations of the adhesion energy from one moir\'{e} site to another \cite{Christian2017}. They are especially strong with metal surfaces like Ni(111) or Ru(0001) in contact with graphene, where the `lock-in' scenario is experimentally the most relevant one. They are much fainter, but non vanishing for Ir(111), even smaller for Cu(111) or Au(111); for these graphene substrates the `lock-in' scenario is expectedly relevant within a limited range of substrate strain. The scenario can be translated into a criterion characterizing the relative in-plane atomic positions of the 2D material and its substrate, at a specific site of the moir\'{e} unit cell. There, the relative atomic positions must be unchanged (and the number of atoms within the unit cell too) during the process that is imposed to the system. In practice, we need to consider different sites depending on the shape of the nanorippling, especially whether it is a 2D cosine (graphene/Ir) or a 2D sine (MoS$_2$/Au). For the former, the site is a moir\'{e} hill, located at the end of the $\vec{u}_\mathrm{m}$ moir\'{e} unit vector \cite{Busse2011}; for the latter, the site is located at the end of $\vec{v}_\mathrm{m}+1/2\vec{u}_\mathrm{m}$ \cite{Sant2020,Silva2022} ($\vec{v}_\mathrm{m}$ being the second moir\'{e} unit vector).

Our calculation of the relative atomic positions requires a 2D integration of distances along a nanorippled surface, a task that we perform numerically, no simple analytical expression being available in the present case. It is hence not possible to directly invert the problem, namely to deduce $\Lambda$, $\Delta$ and $a$ for a given value of the relative atomic positions. Instead we use extended sets of $\lbrace \Lambda,\Delta,a\rbrace$ values (typically, 10$^6$), select those (few 10$^3$) yielding relative atomic positions within a 0.4\% tolerance range, and calculate their total energies, to figure out which of these has the lowest-energy. Strictly speaking, this is not the system's equilibrium configuration, but the most favorable one within the `locked-in' scenario. The process is repeated for all $\varepsilon_\mathrm{s}$ values. Note that for certain  $\varepsilon_\mathrm{s}$ values, a second lowest-energy configuration with close-in energy is found (by close-in, we mean that the configuration is higher by only 10\% or less of the full energy span of the calculated configurations).

\begin{figure}
\begin{center}
\includegraphics[width=8cm]{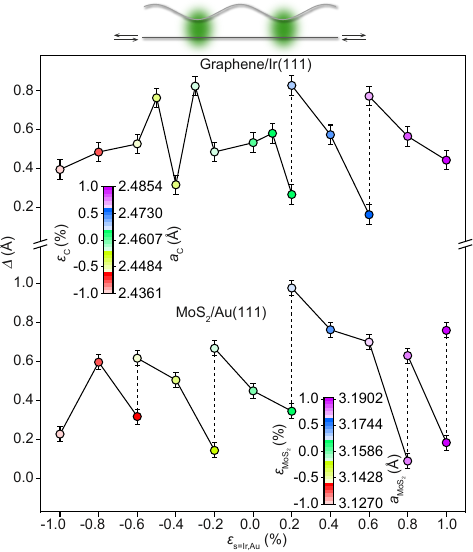}
\caption{\label{fig3} Influence of in-plane biaxial deformation of the substrate lattice, in the absence of interfacial sliding. The variations of $\Delta$ and $a_\mathrm{C,MoS_2}$ in the graphene and MoS$_2$ moir\'{e} lattices are represented as function of the substrate's in-plane deformation $\varepsilon_\mathrm{s}$, with respect to the room-temperature equilibrium in-plane lattice constant. In the calculations, the relative position of the C or S atoms with respect to, respectively, the Ir and Au ones, are kept constant within the valleys when $\varepsilon_\mathrm{s}$ varies.}
\end{center}
\end{figure}

Figure~\ref{fig3} shows the $\Delta$ and $a$ values calculated within this scenario as the substrate strain $\varepsilon_\mathrm{s}$ is varied ($\Lambda$, not shown in the figure, is imposed by the `lock-in' constraint and has only slight variations purely due to the variation of $a_\mathrm{s}$). Not surprisingly, the variations are markedly different from those in Fig.~\ref{fig2}. The most striking observation is the existence of jumps in $\Delta(\varepsilon_\mathrm{s})$. Such jumps also occur in $a(\varepsilon_\mathrm{s})$, but they are much smaller, of about 0.15\% (graphene) and 0.05\% (MoS$_2$).

What is the origin of these jumps? Let us first invoke a purely mechanical argument: a membrane subjected to compressive (tensile) biaxial heteroepitaxial stress is expected to increase (decrease) its $\Delta$ value. The membrane, however, is here also subjected to the interaction with the substrate, and as we have seen earlier $e_\mathrm{ad}$ favours a specific $\Delta$ value. A reduction or increase of $\Delta$ too far from this value, promoted by a mitigation of elastic energy accumulation, will hence become unfavourable at some point. In such case the membrane can increase/decrease its nanorippling if $a$ increases/decreases.

At this point, a few remarks can be made. The abrupt change of $\Delta$ and $a$ appear as discontinuities in the figure (they are marked with vertical dotted lines connecting an upper and a lower branch). For these jumps, the two sets of $\Delta,a$ values correspond to slightly different total energies, and one should expect that repeating the calculations for slightly lower/larger (few 0.01\%) $\varepsilon_\mathrm{s}$ values changes the hierarchy between the two branches, i.e. the jumps are expectedly less abrupt than apparent in the figure. We note that the calculation results presented in Fig.~\ref{fig3} represent a considerable computation time, of several full weeks. Anyway, the observed jump remind a well-known instability of membranes, namely their sudden strain-induced buckling, here controlled by the interaction with the substrate.

\section{Membrane deformation by internal strain}
\label{sec:internalstrain}

In this last section, we turn to the influence of another parameter, which may be seen as the source of what we coin `internal strain'. So far we have assumed that fixed zero-deformation values of the lattice constant within the membrane, $a^0$ (2.462~\AA\; for graphene and 3.167~\AA\; for MoS$_2$,  values measured at 300~K for the flat layers within the bulk version of the materials, see Table~\ref{tab1}). If by some process $a^0$ varies, then the $\varepsilon$ strain within the membrane varies too. The membrane might tend to limit the corresponding elastic energy penalties with changes of $\Lambda$ and $\Delta$. Several processes may lead to $a^0$ variations. Temperature, for instance, can have this effect in relation with phonon anharmonicity \cite{Mounet2005,Zakharchenko2009,Huang2014,Bondarev2018}, and non trivial variations can be expected, possibly altered by the presence of a substrate \cite{Amorim2013}. Illumination can also be relevant, as envisaged so far in 2D materials others than graphene and MoS$_2$, for example via photostriction \cite{Haleoot2017} or light-induced electronic states \cite{Liu2022}.

\begin{figure}
\begin{center}
\includegraphics[width=8cm]{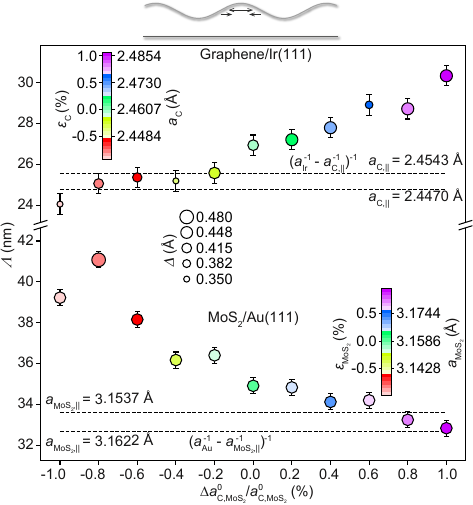}
\caption{\label{fig4} Influence of the zero-deformation value $a^0_\mathrm{C,MoS_2}$ of the lattice constant of the membrane. The variations of $\Lambda$, $\Delta$ and $a_\mathrm{C,MoS_2}$ in the graphene and MoS$_2$ moir\'{e} lattices are represented as function of the variations of  $a^0_\mathrm{C,MoS_2}$, $\Delta a^0_\mathrm{C,MoS_2}/a^0_\mathrm{C,MoS_2}$. The $a_\mathrm{C,MoS_2}$ values are coded with the color of the symbols; the $\Delta$ values are represented with the size of the disk-symbols. Dotted lines show values calculated disregarding the influence of nanorippling, for two sets of room-temperature experimental values of the in-plane projection of the 2D material's lattice constant [values from Refs.~\citenum{Blanc2012} and \citenum{Sant2020} for graphene/Ir(111) and MoS$_2$/Au(111) respectively].}
\end{center}
\end{figure}

Figure~\ref{fig4} shows the calculated $\Lambda,\Delta,a$ variations as a function of $a^0$. There is a global increase of $\Lambda$ as $a^0$ increases in the case of graphene/Ir, and a global decrease for MoS$_2$/Au, at variance with the $\Lambda(\varepsilon_\mathrm{s})$ trend observed in Fig.~\ref{fig2}. This is actually expected, at least qualitatively: here, increasing $a^0$ tends to effectively decreases $\varepsilon$ (strain within the membrane). Although the extent of the $\Lambda$ variations is close to that in Fig.~\ref{fig2}, close inspection reveals several differences. More importantly, the $a$ variations are much different: in both graphene and MoS$_2$, $a$ increases \textit{similarly} to $a^0$ yet with some small differences, i.e. $\varepsilon\sim\Delta a^0/a^0$. This is accommodated in different ways in graphene and MoS$_2$: in the former, $\Lambda$ steadily increases with $\Delta a^0/a^0$, and the membrane adjusts $\Delta$, while in the latter, $\Lambda$ occasionally decreases, and $\Delta$ is constant. This is, once more, a manifestation of $e_\mathrm{ad}$'s strong driving force to select a well-defined value of $\Delta$ in the case of MoS$_2$. Said differently, $\Delta$ is not, for this material and in this range of $\Delta a^0/a^0$ values, a free parameter that the membrane can adjust to minimize its energy.

The comparison between Figs.~\ref{fig2} and \ref{fig4} illustrates that directly changing $a$ (in Fig.~\ref{fig2} changes of $a_\mathrm{s}$ are considered, but as mentioned earlier they are equivalent to changes of $a$) does not have the same effects as changing $a^0$. The reason is that a change of $a^0$ directly translates into an in-plane elastic energy penalty/gain, varying like $\varepsilon^2$, while the direct effect of a change of $a$ is to alter the relative atomic coincidences between the substrate and membrane atoms, leading to an $e_\mathrm{ad}$ penalty/gain (which can then be mitigated with elastic energy changes).

\section{Summary}

Using a model mixing a continuum mechanics description of 2D membranes and an atomistic description accounting for the interaction of these membranes with crystalline substrates, we have explored how epitaxial graphene/Ir(111) and MoS$_2$/Au(111) adapt to different forms of strains, implemented in different situations. We explored the regime of small strains, which seems most relevant experimentally.

First, we considered planar, biaxial strain within the substrate (which is somehow equivalent to a strain within the membranes). In this first situation the membranes were let free to adopt a lowest-energy configuration, their equilibrium configuration, and we found that nanorippling, and the related energy cost, causes a departure from the purely geometrical standard estimate of the moir\'{e} wavelength usually assumed in the literature ($\Lambda=1/|1/a_\parallel-1/a_\mathrm{s}|$). We also found that MoS$_2$ is more strongly influenced by its Au substrate and therefore does not readily change its nanorippling amplitude, while graphene has more freedom in this respect on an Ir substrate. Both materials can also adapt their lattice parameter to manage the substrate biaxial strain.

We next turned to another situation, where the 2D membrane is not free anymore to find its equilibrium state when the substrate is strained, but rather is locked on it, via the moir\'{e}. In this experimentally relevant situation, the membrane is pinned on its substrate at the valleys of the moir\'{e} (where the 2D material is in strongest interaction with its substrate), and the moir\'{e} wavelength is not a free parameter. The membrane experiences a series of transformations, with jumps of its nanorippling amplitude and of its lattice parameter as the substrate strain varies, and overall its lattice constant increases with increasing substrate strain.

Finally, we considered a situation when the zero-deformation value $a^0$ of the lattice constant in the 2D material is varying, and sought for minimum-energy configurations once more. Reminiscent of the effect of substrate strain, a global increase of the moir\'{e} wavelength with $a^0$ is obtained. The 2D material's lattice constant tends to follow, at least to some extent, the $a^0$ variations; the membrane can now change its nanorippling amplitude (in the case of graphene/Ir), and when it cannot, on the contrary needs to adjust its moir\'{e} wavelength, which complexifies the evolution of the latter.

Our analysis could serve as a starting point to quantitatively understand certain mechanical instabilities of on-substrate 2D materials. Strains, induced by temperature for instance, alter the at-equilibrium mechanical state of the membranes. Certain hinderances can prevent the material from adopting this low-energy state, and processes such as the build-up of lateral compressive stress, due to a lattice locking onto the substrate might be relevant in practice. The 2D material itself might change its (internal) mechanical properties under external stimulii (e.g. via photostriction or with a change of temperature \cite{Roldan2011,Bondarev2018,Sajadi2018}, in relation with flexural modes that are strongly influenced by the presence of the substrate \cite{Amorim2013}). Carefully studying the relative importance of these different effects could allow to quantitatively understand the formation, density and morphology of wrinkles in epitaxial 2D materials in the future. Intuitively at least, one expects wrinkles to locally introduce what resembles open boundary conditions for the moir\'{e} lattice, acting as more-or-less free-to-move edges letting the moir\'{e} lattice expand or un-ripple in response to planar stress.

After first works focused on the influence of the interaction potential between the 2D membrane and their substrate \cite{Artaud2022}, the present work analyzed the role of strains, external ones from the substrate and internal ones within the membrane itself. We so far disregarded the effect of the twist angle between the two materials, which would require a dedicated study. Considering substrates with small lattice mismatch with the 2D material, e.g. Cu(111) for graphene, would bring the interesting question of the reconstruction of the moi\'{e} lattice, possibly producing a spatially non-uniform lattice constant as observed in other large-$\Lambda$ systems \cite{Woods2014,Gadelha2021}, which can be treated using the theoretical framework we used. Systematically investigating the influence of each of the mechanical constants of the membranes, which can vary with temperature for example, is another possible direction for future investigations. Our results can serve as a basis to understand and conceive experimental endeavours toward the engineering of moir\'{e} patterns -- their wavelength, nanorippling amplitude --, for instance via mechanical stress, temperature, light, or the presence of imperfections in the substrate (e.g. step edges, dislocations emerging at the surface).

\section{acknowledgements}
We thank Vincent T. Renard and Nicolae Atodiresei for fruitful discussions.


%

\end{document}